\documentclass{elsart}
\usepackage{graphicx}
\usepackage{amsmath}
\usepackage{amssymb}
\def\be{\begin{equation}}
\def\ee{\end{equation}}

\begin{document}

\begin{frontmatter}
\title{Accuracy and Robustness of Clustering Algorithms for Small-Size Applications in Bioinformatics}

\author[address1]{Pamela Minicozzi}
,
\author[address1]{Fabio Rapallo}
,
\author[address1,thank1]{Enrico Scalas}
and
\author[address2]{Francesco Dondero}

\address[address1]{Department of Advanced Sciences and Technology,\\
Universit\`a degli Studi del Piemonte Orientale,
\\ via Bellini 25g, 15100 Alessandria, Italy}

\address[address2]{Department of Life and Environmental Science, \\
Universit\`a degli Studi del Piemonte Orientale, \\
via Bellini 25g, 15100 Alessandria, Italy}

\thanks[thank1]{
Corresponding author. \\
E-mail: enrico.scalas@mfn.unipmn.it}

\begin{abstract}
The performance (accuracy and robustness) of several clustering
algorithms is studied for linearly dependent random variables in
the presence of noise. It turns out that the error percentage
quickly increases when the number of observations is less than the
number of variables. This situation is common situation in
experiments with DNA microarrays. Moreover, an {\it a posteriori}
criterion to choose between two discordant clustering algorithm is
presented.
\end{abstract}


\end{frontmatter}

%[main text]
\section{Introduction} \label{introsect}

Multivariate statistical techniques are an essential tool in many
fields of applied science, including Physics, Computer Science,
Biology, Medicine, Finance and Economics. In recent years, thanks
to the availability of powerful computing tools, such methods have
received increasing attention. Among them, {\em cluster analysis}
or {\em clustering} is used for partitioning available data into
groups when prior information is not available or limited. A set
of $n$ objects can be allocated into $g$ categories in
$\dbinom{n+g-1}{g-1} = \dbinom{n+g-1}{n}$ different ways. This
number soon becomes very large so that a study by direct
enumeration of all possible clusters is no longer tractable. With
$n=20$ objects and $g=10$ categories, one already has more than
ten million possible clusters.

Clustering defines the class of {\em unsupervised} classification
methods \cite{xu|wunsch:05}. This means that clustering separates
a finite data set into a finite number of ``natural'' categories,
where the word {\em natural} has to be specified according to some
measure of closeness between data. Unsupervised classification is
opposed to {\em supervised} classification, where one looks for an
accurate characterization of samples generated from some
probability distribution with some {\em a priori} knowledge.

This paper was originally motivated by microarray data analysis.
Indeed, Cluster analysis is becoming a major tool in
bioinformatics \cite{bendor|shamir|yakhini:99}, \cite{miller:99},
\cite{yeung|haynor|ruzzo:01}, and \cite{sasson|linial|linial:02}
and it is widely applied in microarray data analyses: its use is
rapidly growing in a wide range of microarray-related problems
(see \cite{baldi|long:01} and
\cite{moreau|smet|thijs|marchal|moor:02}).

In a typical microarray experiment, the expression of several
thousands of genes is compared in different experimental
conditions. The expression is given by the variable
\begin{equation*}
x = \log_2 \left( \frac{I_R}{I_G} \right),
\end{equation*}
where $I_R$ is the (red) fluorescent intensity coming from
reference spots and $I_G$ is the (green) fluorescent intensity
coming from treated spots.

After proper normalization and filtering, only a few or at most
hundreds of genes result as significantly differentially
expressed. This means that $x$ is significantly different from
zero. Then, it is useful to apply clustering algorithms in order
to detect common patterns of differentially expressed genes. Small
groups of genes can be obtained without considering a priori the
expression levels, but from a functional analysis through
dedicated bioinformatics tools, such as the Gene Ontology
annotation \cite{geneontology:00}. The Gene Ontology is a
controlled vocabulary to describe gene and gene product
attributes, virtually, in any organism. The Gene Ontology is
structured as directed acyclic graphs in which terms are
classified in levels and linked through a parent/child
relationship. This feature permits to select genes sharing common
terms into relative large or small groups depending on the level
one is looking at.

From the statistical viewpoint, it is interesting to investigate
the behavior of clustering algorithms when the sample size is
small. In fact, many statistical procedures dramatically lose
accuracy and robustness for small sample sizes.

Moreover, in microarray experiments, data are affected by noise
due to measurement errors. Although simple and visually appealing,
the performances of clustering algorithms are in general sensitive
to noise. Thus, a crucial question is to study their robustness to
noise. Some papers in this direction are \cite{kerr|churchill:01},
\cite{bickel:03} and \cite{datta|datta:06}. The central subject of
these works is to perform Monte Carlo simulations to evaluate the
behavior of the clustering algorithms. In many cases, the authors
define numerical indices to capture the robustness of a clustering
algorithm, but such indices are often criticized in subsequent
papers.

Having explained our motivations, from now on, we will consider a
rather general clustering problem. To illustrate this problem on a
synthetic example, let us consider the data in Table \ref{tabex}
with $6$ genes and $6$ experimental conditions $X_1, \ldots, X_6$.

{\tt Table \ref{tabex} approx here.}

We applied two different clustering algorithms on the column of
the matrix in Table \ref{tabex}. Requiring a final partition with
$3$ clusters, we have the following results:
\begin{itemize}
\item with the single-linkage technique, the final partition is
$\{X_1 , X_2, X_5, X_6\}$, $\{X_3 \}$, $\{X_4\}$;

\item with a non-hierarchical technique (PAM), the final partition
is $\{X_1, X_2\}$, $\{X_3, X_4\}$, $\{X_5, X_6 \}$.
\end{itemize}

The clustering techniques will be presented in section
\ref{dessect} with some details. Therefore, a question naturally
arises. We have to evaluate what final partition is more likely
or, in other words, what algorithm is more accurate in our special
case.

In order to answer this question, we can follow two approaches:
\begin{itemize}
\item we study the accuracy of the clustering algorithms in
several known configurations;

\item we make use of the Bayes formula to measure the accuracy of
each algorithm.
\end{itemize}
We will see in the next section that both approaches produce
useful information to address this problem.

We study the accuracy and robustness of various clustering
algorithms when the distribution of the variables becomes
heavy-tailed and for different sample sizes. By accuracy we mean
the insensitiveness to noise, while robustness stands for
insensitiveness to heavy tails. In particular, we concentrate on
small sample sizes. This has been done with a Monte Carlo study
where the simple assumption of linearly correlated random variable
is used. Moreover, we show how to use the Bayes formula to give an
{\it a posteriori} measure of accuracy for two competing
clustering results. We apply this technique to the real data
example presented earlier.

The material is organized as follows. In section \ref{algosect},
some relevant clustering algorithms are described, while in
section \ref{dessect} we present the design of the simulation
study and we discuss the choice of the parameters. The results are
summarized in section \ref{ressect}. In section \ref{aposteriori}
we make use of the Bayes rule and of the Monte Carlo algorithms to
give a measure of the accuracy when two algorithms produce
different partitions and we present a  numerical example. Finally,
section \ref{finremsect} is devoted to a discussion of the major
findings and of the pointers to future research.

\section{The clustering algorithms} \label{algosect}

In this section we briefly review the algorithms we have compared
in our simulation study.

There are perhaps uncountable many algorithms for doing cluster
analysis. However, they belong to one of the following categories:
\begin{itemize}
\item agglomerative hierarchical methods;

\item divisive hierarchical methods;

\item non-hierarchical methods.
\end{itemize}
Therefore, we have compared three algorithms, one for each of the
three categories: hierarchical single-linkage, {DI}visive
{ANA}lysis (DIANA) and Partitioning Around Medoids (PAM) where a
medoid can be defined as that object of a cluster, whose average
dissimilarity to all the objects in the cluster is minimal. For
details on these methods the reader can refer to
\cite{everitt:01}, \cite{kaufman|rousseeuw:05} and
\cite{everitt:05}. We have chosen single-linkage, DIANA and PAM
because of their sensitivity to outliers, see for instance
\cite{everitt:01}, and thus they are appropriate to study the
robustness to heavy tailed distributions.

The hierarchical algorithm begins by assigning each point to its
own cluster (i.e., each group contains just one point). At each
stage the distances between clusters are computed. The
single-linkage method defines the distance between two clusters as
the minimum distance between the points in one cluster and the
points in the other cluster. Once the distances are computed, the
two nearest clusters are merged together. These steps are repeated
until all points are clustered into a single cluster of size $n$.
It is known that single-linkage is particulary appropriate to
detect outliers, i.e., when a cluster contains only one point.

On the other hand, the DIANA algorithm starts with a single
cluster of size $n$. At each step, the algorithm selects the
cluster with the maximum diameter. Then, it creates a new cluster
containing the point with the largest average dissimilarity with
respect to the other points and assigns to this new cluster the
points closer to the new cluster than to the old one. The
algorithm proceeds until the $n$ subjects belong to $n$ distinct
clusters.

In the PAM algorithm, the user must decide {\em a priori} the
final number $k$ of clusters. First, the algorithm randomly splits
the data into $k$ clusters. Then, it computes the medoids for each
cluster and each point is assigned to the closest medoid. The
medoids are recalculated every time an observation is added to the
cluster. All steps continue until no points has to be moved or
some stability criteria is satisfied.

As a distance between two variables, say $X$ and $Y$, we have used
the Pearson correlation distance
\begin{equation}
d(X,Y) = \sqrt{2(1-\rho(X,Y))} \, ,
\end{equation}
where $\rho$ denotes the Pearson's correlation coefficient. Notice
that the range of $d(X,Y)$ is the closed interval $[0,2]$,
$d(X,Y)=2$ if and only if $\rho(X,Y)=-1$, and $d(X,Y)=0$ if and
only if $\rho(X,Y)=1$. Therefore, that distance detects pairs of
variables with strong positive linear dependence.

An useful and easy reference for all methods can be found in the
user's manual of the {\tt R}-package {\tt cluster}, see
\cite{r-cluster}, which also presents relevant numerical issues
and a number of examples based on real data.

\section{Study design} \label{dessect}

Every simulation study in the field of cluster analysis has an
impressive number of parameters. Thus, we have to restrict our
study to special settings.

Based on the discussion presented in the introduction, we have
considered data sets with $p=6$ variables $X_1, \ldots, X_6$ and
final partitions with $3$ clusters. Apart from permutations, there
are $3$ possible patterns in that situations:
\begin{itemize}
\item pattern $1$: $S_1=\{X_1,X_2\}$, $S_2=\{X_3,X_4\}$,
$S_3=\{X_5,X_6\}$;

\item pattern $2$: $S_1=\{X_1,X_2,X_3\}$, $S_2=\{X_4,X_5\}$,
$S_3=\{X_6\}$;

\item pattern $3$: $S_1=\{X_1,X_2,X_3,X_4\}$, $S_2=\{X_5\}$,
$S_3=\{X_6\}$.
\end{itemize}

We assume that data are arranged in a matrix with $6$ columns
(variables) and $n$ rows (objects). We have considered different
sample sizes:
\begin{itemize}
\item $n=10$, i.e. a case where $n>p$;

\item $n=6$, i.e. the special case $n=p$;

\item $n=4$, i.e. a case where $n<p$.
\end{itemize}

We apply the clustering algorithm in order to detect linear
relationships in the set of variables. When two variables, say
$X_i$ and $X_j$, have perfect linear dependence, there is a linear
function
\begin{equation} \label{linrel}
X_j= a + bX_i
\end{equation}
which holds exactly for appropriate coefficients $a$ and $b$. As
mentioned in the introduction, the experimental data present a non
negligible noise. Therefore, the linear relationship in Equation
\ref{linrel} does not hold exactly, but Eq. \eqref{linrel} assumes
the form
\begin{equation} \label{linrelappr}
X_j= a + bX_i + \varepsilon_{ij} \, ,
\end{equation}
where $\varepsilon_{ij}$ is assumed to be a random variable with
mean $0$ and variance $\sigma^2_{ij}$. The additional term
$\varepsilon_{ij}$ in Eq. \eqref{linrelappr} represents the noise.
We restrict our study to linear relationships among variables
because it is difficult to detect more complex relationships with
small samples. Moreover, the distance we use is based on the
correlation coefficient, which is a measure of linear
dependencies. Nevertheless, in principle, one can modify the
distance definition and consider relationships with different
functional forms.

To study the robustness of the algorithms, we have used two
distribution functions to generate the variables of the data set:
\begin{itemize}
\item the standard normal distribution;

\item the (standardized) Student's $t$ distribution with $3$
degrees of freedom.
\end{itemize}
The Student's $t$ with $3$ degrees of freedom is heavy-tailed but
its variance is finite. We have used a standardized Student's $t$
in order to allow the comparisons with the standard normal
distribution.

Moreover, to study the accuracy of the algorithms, we have
considered increasing variances of the noise. The standard
deviation of the $\varepsilon_{ij}$ terms in Eq.
\eqref{linrelappr} has been taken from $1/30$ to $1$ for each
numerical experiment. This means that at the final stage (when the
standard deviation equals to $1$) the noise has the same variance
as the independent variable.

In our experimental design we consider the clustering of the
variables. However, once the distance matrix is computed, the
algorithms also work to cluster observations.

In formulae, the data sets for the standard normal variables and
the first pattern are generated as follows:
\begin{itemize}
\item $X_1 \sim {\mathcal N}(0,1)$, $X_2=X_1 + \varepsilon_{12}$;

\item $X_3 \sim {\mathcal N}(0,1)$, $X_4=X_3 + \varepsilon_{34}$;

\item $X_5 \sim {\mathcal N}(0,1)$, $X_6=X_5 + \varepsilon_{56}$.
\end{itemize}
where the error columns $\varepsilon _{ij}$ are normally
distributed, $\varepsilon _{ij} \sim {\mathcal
N}(0,\sigma^2_{ij})$. The formulae for all patterns are in Table
\ref{table:structure}. Notice that the correlation coefficient is
insensitive with respect to linear transformations of the
variables with positive slopes. Therefore, in our study it is
sufficient to consider the special case of linear transformations
with coefficients $a=1$ and $b=0$.

{\tt Table \ref{table:structure} approx here.}

We have considered two different approaches. The first one uses a
purely parametric Monte Carlo algorithm. At each iteration, our
algorithm generates the independent variables and the errors, and
computes the dependent variables. The standard deviations of the
$\varepsilon_{ij}$ represent the magnitude of the noise. As said
before, the $\varepsilon_{ij}$ have the same distribution as the
independent variables, except for the parameters. Once the data
set is completed, the algorithm applies the three clustering
techniques and checks the correctness of the results. By virtue of
the small number of variables the correctness is evaluated in
binary form (correct/uncorrect). A second approach is a
non-parametric one, where the standard deviations of the noise
terms are estimated from the observed data set. As this second
approach produces results very close to the parametric case, we do
not present the non-parametric results in the next section.

For each parameter configuration, the simulation study is based on
$B=10,000$ iterations and the results are displayed as a plot of
the error rate ($f$) versus the nuisance parameter $\sigma$. The
most informative plots are presented and discussed in the next
section. The simulation study and the presentation of the results
are both implemented in {\tt R} using the additional package {\tt
cluster}.

\section{Results} \label{ressect}

In this section we present the main results of our simulation
study. In all figures, we plot the percentage of errors ($f$)
versus the standard deviation of the noise ($\sigma$).

\begin{figure}
\begin{center}
\includegraphics[width=12cm]{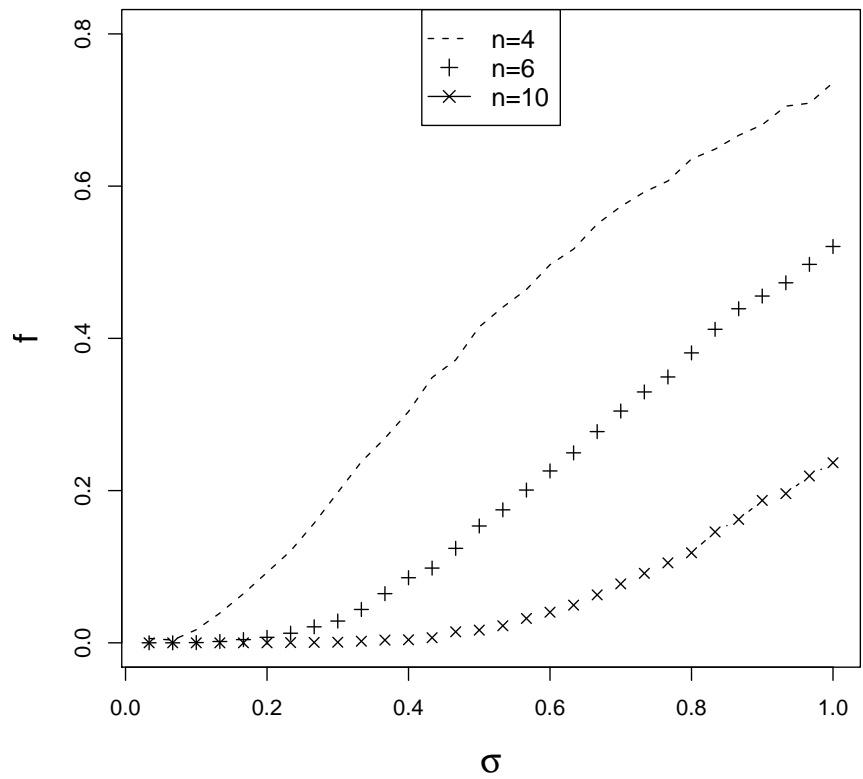}
\caption[]{Percentage of errors (f) versus the noise standard
error $\sigma$ for the DIANA algorithm with normal distribution
and pattern $1$.}\label{diana_n_p1}
\end{center}
\end{figure}

\begin{figure}
\begin{center}
\includegraphics[width=12cm]{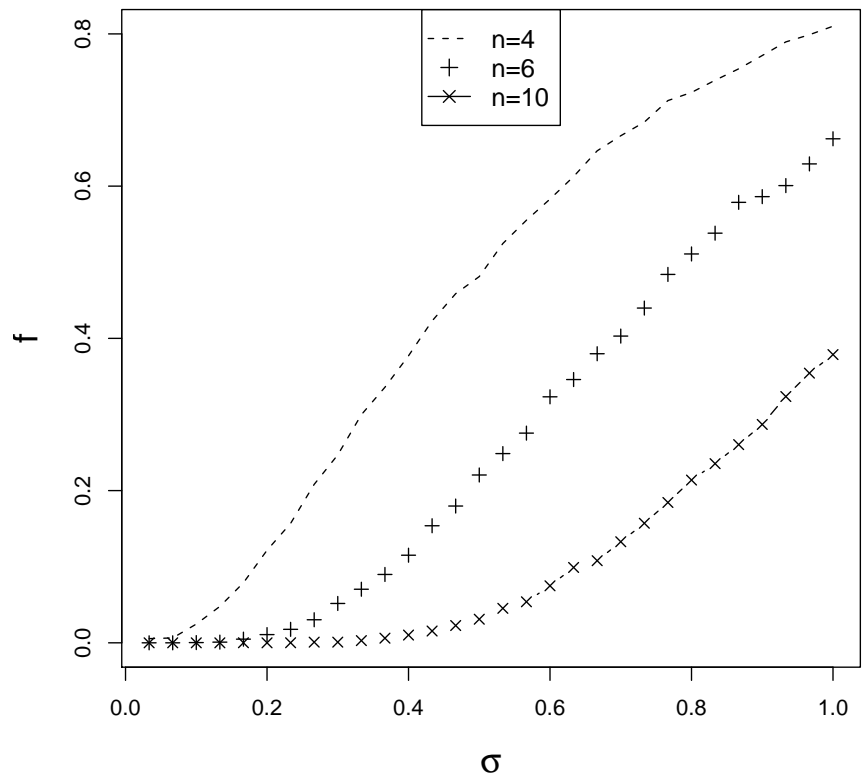}
\caption[]{Percentage of errors (f) versus the noise standard
error $\sigma$ for the single-linkage algorithm with normal
distribution and pattern $1$.}\label{single_n_p1}
\end{center}
\end{figure}

{\tt Figures \ref{diana_n_p1} and \ref{single_n_p1} approx here.}

The plot in Figure \ref{diana_n_p1} shows the behavior of the
DIANA method when variables are normally distributed and clustered
according to the first pattern. When the number of observations
decreases, the error percentage dramatically increases. For
instance, using the DIANA algorithm, one gets an error rate of
$10\%$ for $\sigma \simeq 0.2$ when $n=4$, for $\sigma \simeq
0.45$ when $n=6$, for $\sigma \simeq 0.8$ when $n=10$, see Figure
\ref{diana_n_p1}. A similar behavior is shown in Figure
\ref{single_n_p1}, where we applied the single-linkage algorithm
to normally distributed variables, always according to the first
pattern. This is also true for the PAM algorithm, not displayed.

\begin{figure}
\begin{center}
\includegraphics[width=12cm]{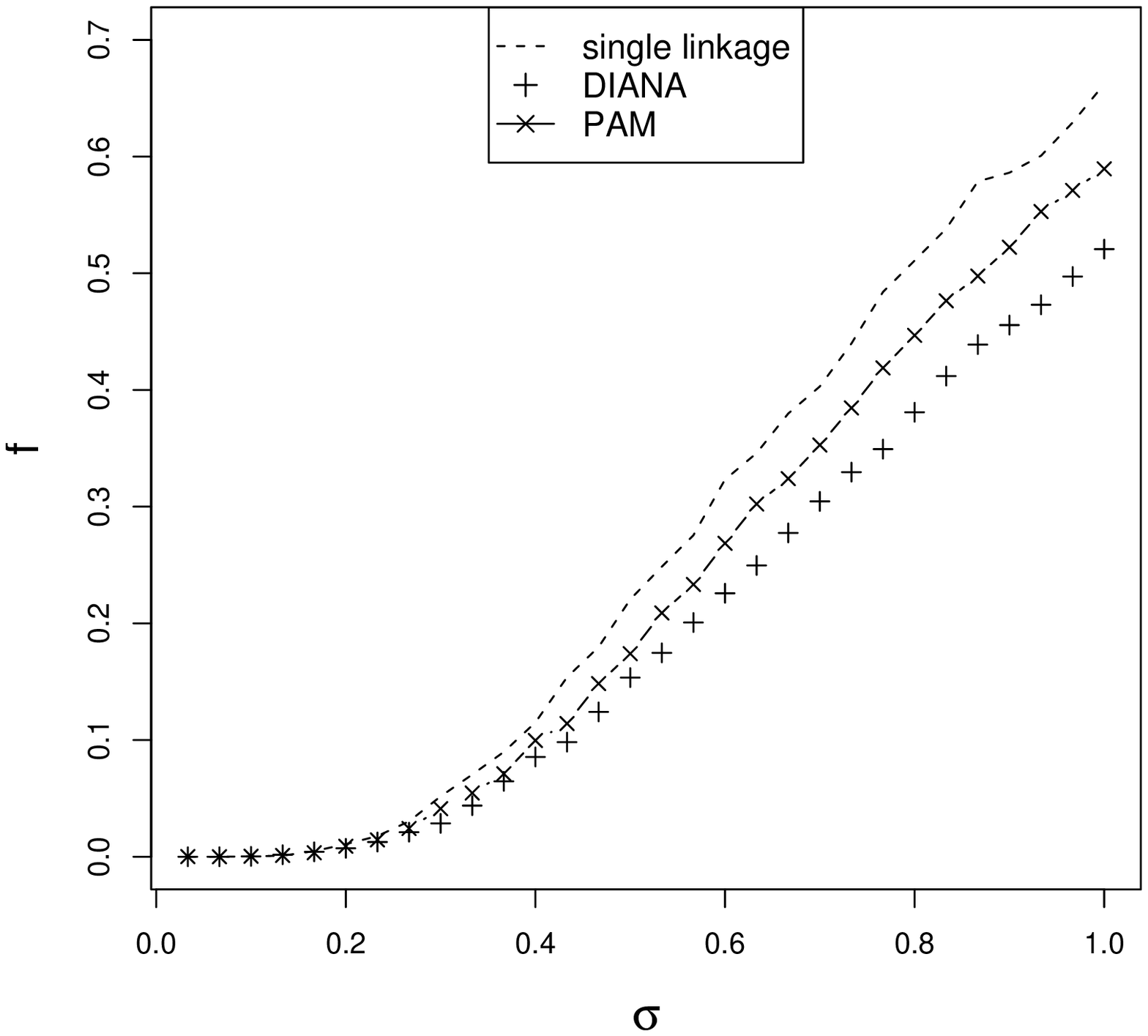}
\caption[]{Percentage of errors (f) versus the noise standard
error $\sigma$ for sample size $n=6$ with normal distribution and
pattern $1$.} \label{n6_n_p1}
\end{center}
\end{figure}

\begin{figure}
\begin{center}
\includegraphics[width=12cm]{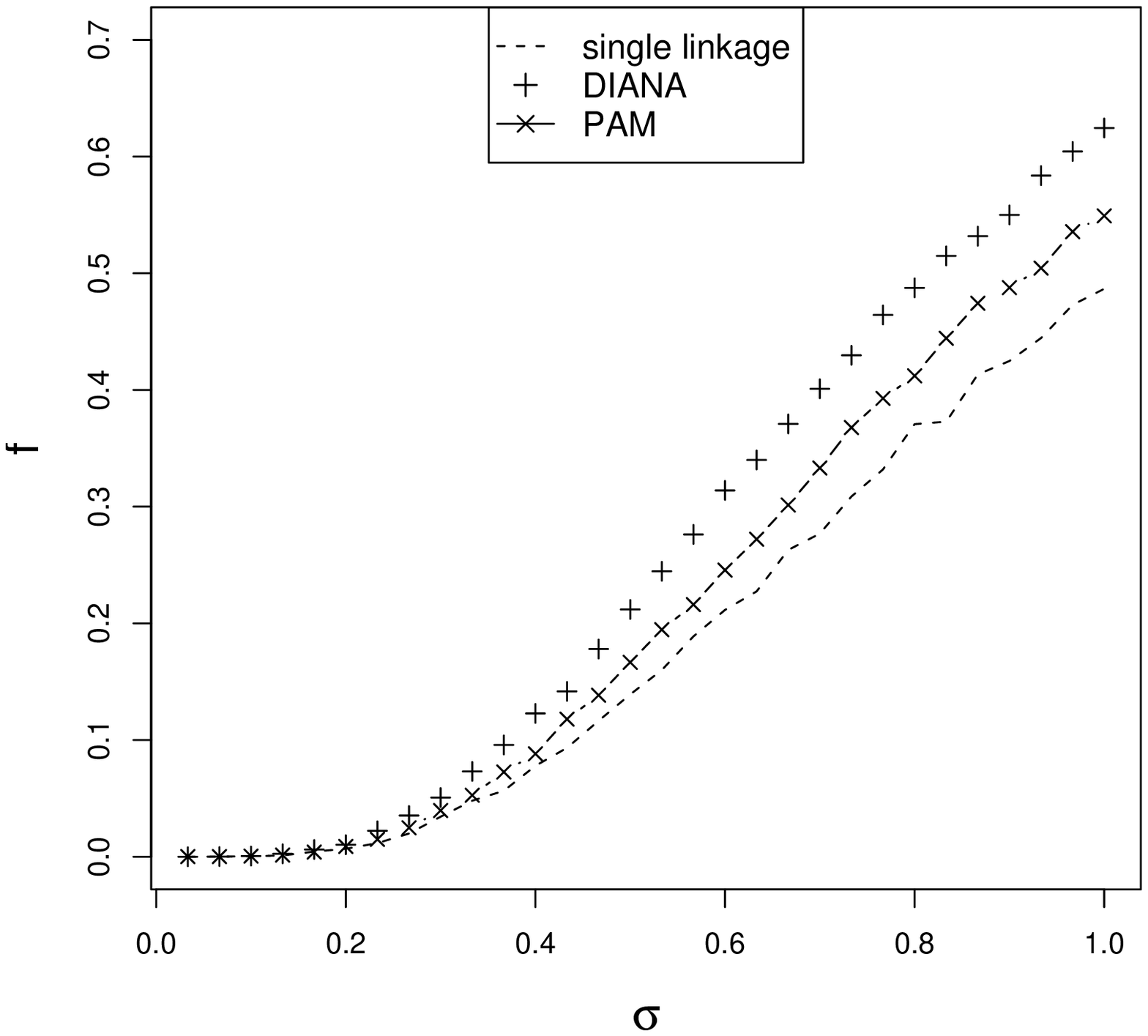}
\caption[]{Percentage of errors (f) versus the noise standard
error $\sigma$ for sample size $n=6$ with normal distribution and
pattern $3$.}\label{n6_n_p3}
\end{center}
\end{figure}

{\tt Figures \ref{n6_n_p1} and \ref{n6_n_p3} approx here.}

In Figure \ref{n6_n_p1} the three methods are compared while
working on normally distributed data with sample size $n=6$ and
pattern $1$. Notice that the hierarchical clustering method gives
worse results than the other algorithms when the variables are
grouped according to the first pattern. This is in agreement with
the known properties of the single-linkage method, which presents
the best performance in the presence of singletons. On the other
hand, in Figure \ref{n6_n_p3}, where we used pattern $3$ with $2$
singletons, the hierarchical clustering method shows the best
performance.

\begin{figure}
\begin{center}
\includegraphics[width=12cm]{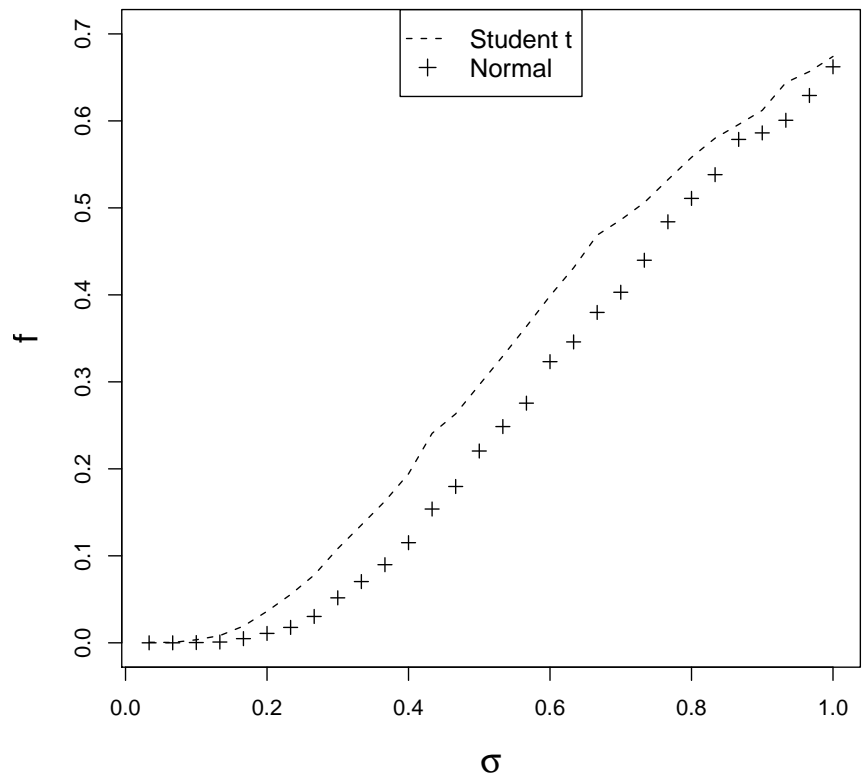}
\caption[]{Percentage of errors (f) versus the noise standard
error $\sigma$ for the single-linkage algorithm with sample size
$n=6$ and pattern $1$.}\label{n6_single_p1}
\end{center}
\end{figure}

\begin{figure}
\begin{center}
\includegraphics[width=12cm]{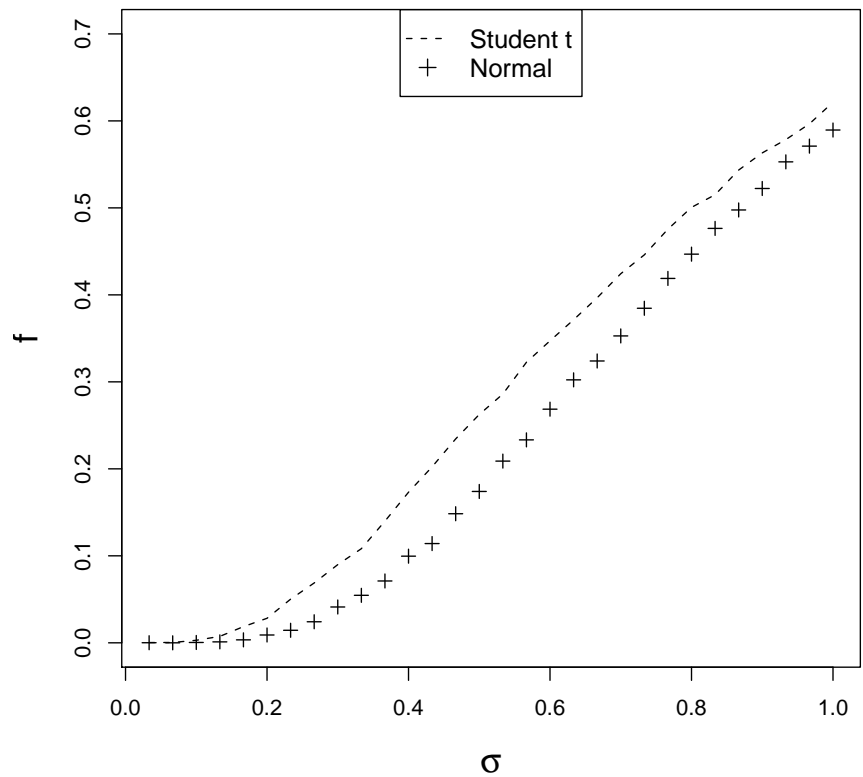}
\caption[]{Percentage of errors (f) versus the noise standard
error $\sigma$ for the PAM algorithm with sample size $n=6$ and
pattern $1$.}\label{n6_PAM_p1}
\end{center}
\end{figure}

{\tt Figures \ref{n6_single_p1} and \ref{n6_PAM_p1} approx here.}

In Figure \ref{n6_single_p1} and \ref{n6_PAM_p1} the
single-linkage method and the PAM method are compared while
working on a sample size $n=6$ and pattern $1$. We can observe
that in both cases the error rate is slightly lower for the normal
distribution and larger for the Student distribution. Therefore,
clustering methods do lose robustness when applied to heavy tailed
distributions, but this loss is not very large. There is a wide
literature on the effects of heavy tails on regression analyses.
Here, we limit ourselves to a simple graphical summary. The
interested reader can find analytical and numerical estimates of
such effects in e.g. \cite{adler|feldman|taqqu:95}, where further
literature pointers are available.

\section{Evaluation of the accuracy of competing algorithms} \label{aposteriori}

Suppose now that two clustering algorithms produce two different
final partitions. In order to choose between the two competing
results, we need a measure for the accuracy of the two algorithms.

Suppose that with a first algorithm $A_1$, we observe the
partition $c_1$, while with a second algorithm $A_2$ we observe
the partition $c_2$. We denote by $C_1$ and $C_2$ the
corresponding theoretical patterns.

The comparison of the two algorithms in terms of accuracy can be
performed by computing the conditional probabilities ${\mathbb
P}(C_1 | c_1)$ for $A_1$ and ${\mathbb P}(C_2 | c_2)$ for $A_2$.
In other words, we compute the {\it a posteriori} probabilities
that the theoretical models are correct. As a criterion, we
suggest to select the algorithm producing the highest value of
${\mathbb P}(C_i | c_i)$.

Following the Bayes' rule the first probability can be computed
under $A_1$ as
\begin{equation*}
{\mathbb P}(C_1 | c_1) = \frac {{\mathbb P}(c_1 | C_1) {\mathbb
P}(C_1)} {{\mathbb P}(c_1 | C_1) {\mathbb P}(C_1) + {\mathbb
P}(c_1 | C_2) {\mathbb P}(C_2)}
\end{equation*}
and similarly for ${\mathbb P}(C_2 | c_2)$ under $A_2$.

The conditional probabilities ${\mathbb P}(c_i | C_j)$, $i,j=1,2$
can be approximated through a simple Monte Carlo simulation as
described in section \ref{dessect}, by setting the noise variance
equal to the estimated variance of the residuals after the least
squares approximation.

The probabilities ${\mathbb P}(C_1)$ and ${\mathbb P}(C_2)$ are
the {\it a priori} probabilities of the two patterns. In our
example, we have no prior knowledge on the behavior of the genes
and thus we set non-informative {\it a priori} probabilities
equals to $1/2$ each.

Notice that the conditional probabilities ${\mathbb P}(C_i | c_i)$
can be viewed as the ``likelihoods'' of the patterns $C_i$.
However, it should be noted that the term likelihood is improper
as the two probabilities are computed using different algorithms.

Now, we apply the previous formula to the data set in Table
\ref{tabex} presented in the introduction. As mentioned there, two
algorithms produce competing results. In particular:
\begin{itemize}
\item with the single-linkage algorithm, the final partition is
\begin{equation*}
c_1=\{\{X_1 , X_2, X_5, X_6\}, \{X_3 \}, \{X_4\} \} \, ;
\end{equation*}
\item with the PAM algorithm, the final partition is
\begin{equation*}
c_2=\{\{X_1, X_2\}, \{X_3, X_4\}, \{X_5, X_6 \} \} \, .
\end{equation*}
\end{itemize}
To evaluate what partition is more reliable, we consider the
corresponding theoretical patterns $C_1$ and $C_2$ and we
approximate the conditional probabilities ${\mathbb P}(C_1 | c_1)$
for the single-linkage algorithm and ${\mathbb P}(C_2 | c_2)$ for
the PAM algorithm. The results are:
\begin{itemize}
\item ${\mathbb P}(C_1 | c_1) = 0.9696$;

\item ${\mathbb P}(C_2 | c_2) = 0.9945$.
\end{itemize}
Therefore, we are more confident in the result $c_2$ obtained with
the PAM algorithm than in the result $c_1$ obtained with the
single-linkage. Indeed, the dataset in Table \ref{tabex} was
generated under the theoretical pattern $C_2$, which has the
largest conditional probability.

\section{Final remarks and future work} \label{finremsect}

\subsection{Summary}

In this paper, we discussed the performance (accuracy and
robustness) of several clustering algorithms. Assuming linear
dependence between random variables, we checked to what extent
clustering algorithms are able to identify given clusters when
noise is introduced. We also investigated the effect of heavy
tails, by comparing normally distributed residuals with Student
$t$ distributed residuals. Finally, we suggested a simple way to
discriminate between different results given by competing
clustering methods, based on Bayes' formula.

Some of our results are not surprising. For instance, in the
presence of singletons, the hierarchical single-linkage method is
better performing, as expected. Moreover, the algorithm
performance is better when residuals are not heavy-tailed.

However, some results deserve particular attention:
\begin{enumerate}
\item The error percentage quickly increases when the number of
observations is less than the number of variables, a common
situation in many experiments with DNA microarrays;

\item Also for $n=10$ (i.e. when the number of observations
exceeds the number of variables), for $\sigma$ greater than about
$0.8$, the noise produces an error rate larger than $20\%$ in
almost all settings. This means that clustering results should be
considered with great care;

\item Simulation studies show that the influence of heavy tails is
not as significant as the effect of noise (see Figures $3$ to $6$
for comparison).
\end{enumerate}

Moreover, these results are corroborated by a non-parametric study
where the noise level is estimated based on $5,000$ Monte Carlo
data sets. All the trends discussed in section \ref{dessect} can
be reproduced within the non-parametric approach.

\subsection{Outlook}

In future papers, these analyses will be extended in three
directions. A paper will be devoted to a detailed study on the
accuracy of clustering methods as a function of the number of
observations and the number of variables, focusing on small
numbers of both variables and observations. Another paper will
explore an alternative clustering method based on random matrix
theory that has been recently proposed, see \cite{luoetal:06}.
Also in this case, we will try to assess the accuracy and
robustness of the method. A third study will concern the situation
with many variables and few observations, which is usual in
microarray experiments. In this case, a binary evaluation of the
correctness for an algorithm is too strict. For example, two trees
with hundreds of nodes and differing only for a few nodes can
essentially be considered as equivalent.

\section*{Acknowledgements}
This paper is part of a research project devoted to understanding
the effect of noise in unsupervised clustering methods and was
also motivated by measurements carried out using microarrays on
different environmental relevant model species exposed to adverse
conditions, see \cite{donderoetal:06a}, \cite{donderoetal:06b} and
\cite{donderoetal:06c}. This work has been supported by two
Italian grants: MIUR PRIN 2006 (Minicozzi, Scalas, Dondero) and
UPO ``Ricerca Locale'' 2007 (Rapallo).

\bibliographystyle{elsart-num}
\bibliography{tutto3}

\newpage

\section*{Tables and figures}

\newpage

\begin{table}
\begin{center}
\begin{tabular}{c|cccccc}
& $X_1$ & $X_2$ & $X_3$ & $X_4$ & $X_5$ & $X_6$ \\ \hline

$G_1$ & $-0.440$ & $0.563$ & $-0.452$ & $-1.155$ & $1.125$ &
$1.162$ \\
$G_2$ & $-0.531$ & $-0.785$ & $-0.340$ & $-0.793$ & $0.682$ &
$1.003$ \\
$G_3$ & $0.613$ & $1.310$ & $-1.582$ & $-2.209$ & $1.442$ &
$1.966$ \\
$G_4$ & $-0.912$ & $-1.765$ & $-0.491$ & $-0.796$ & $-1.520$ &
$-1.820$ \\
$G_5$ & $1.743$ & $2.185$ & $-1.480$ & $0.003$ & $1.010$ & $1.216$
\\
$G_6$ & $0.422$ & $0.072$ & $1.604$ & $1.136$ & $-0.064$ & $0.238$
\end{tabular}
\end{center}
\caption[]{A $6 \times 6$ matrix from a simulated microarray
experiment. The rows $G_1, \ldots, G_6$ denote the genes; the
columns $X_1 , \ldots, X_6$ denote the tissues.}\label{tabex}
\end{table}

%$G_1$ & $-0.604$ & $0.211$ & $0.004$ & $-0.461$  & $0.622$ &
%$0.405$ \\
%$G_2$ & $-0.697$ & $-0.734$ & $0.101$ & $-0.140$ & $0.217$ &
%$0.285$ \\
%$G_3$ & $0.476$ & $0.735$ & $-0.981$ & $-1.397$ & $0.913$ &
%$1.016$ \\
%$G_4$ & $-1.087$ & $-1.426$ & $-0.030$ & $-0.142$ & $-1.802$ &
%$-1.855$ \\
%$G_5$ & $1.632$ & $1.350$ & $-0.892$ & $0.567$ & $0.517$ & $0.446$
%\\
%$G_6$ & $0.279$ & $-0.134$ & $1.797$ & $1.573$ & $-0.468$ &
%$-0.295$

\newpage

\begin{table}
\begin{center}
\begin{tabular}{c|c|c|c} \hline\hline
 Pattern $1$ &   $(1,2)$  &  $(3,4)$ & $(5,6)$  \\   \hline
&   $X_1 \sim {\mathcal N}(0,1)$   &   $X_3 \sim {\mathcal
N}(0,1)$ & $X_5 \sim {\mathcal N}(0,1)$\\     \hline &  $X_2=X_1 +
\varepsilon _{12}$    &   $X_4=X_3 + \varepsilon _{34}$ & $X_6=X_1
+ \varepsilon _{56}$\\ \hline\hline Pattern $2$ & $(1,2,3)$  &
$(4,5)$ & $(6)$  \\  \hline
& $X_1 \sim {\mathcal N}(0,1)$   &   $X_4 \sim {\mathcal N}(0,1)$ & $X_6 \sim {\mathcal N}(0,1)$\\
    \hline
&  $X_2=X_1 + \varepsilon _{12}$    &   $X_5=X_4 + \varepsilon _{45}$ & $-$\\
    \hline
&  $X_3=X_1 + \varepsilon _{13}$    &   $-$ & $-$\\
    \hline\hline
     Pattern $3$ &   (1,2,3,4)  &  (5) & (6)\\
    \hline
&  $X_1 \sim {\mathcal N}(0,1)$   &   $X_5 \sim {\mathcal N}(0,1)$ & $X_6 \sim {\mathcal N}(0,1)$\\
\hline & $X_2=X_1 + \varepsilon _{12}$    &   $-$ & $-$\\ \hline
& $X_3=X_1 + \varepsilon _{13}$    &   $-$ & $-$\\
    \hline
&  $X_4=X_1 + \varepsilon _{14}$    &   $-$ & $-$\\
    \hline
\end{tabular}
\end{center}
\caption[]{Model equations for the normal
case.}\label{table:structure}
\end{table}

\newpage

\newpage

\newpage

\newpage

\newpage

\newpage

\end{document}